%% file: main.tex
\documentclass[acmsmall,nonacm]{acmart}
\input{dlabmacros}
\usepackage{booktabs}
\usepackage{subcaption}
\AtBeginDocument{%
  \providecommand\BibTeX{{%
    \normalfont B\kern-0.5em{\scshape i\kern-0.25em b}\kern-0.8em\TeX}}}

\setcopyright{none}

\usepackage{bbold}

\begin{document}

\title{Do Platform Migrations Compromise Content Moderation? Evidence from r/The\_Donald and r/Incels}

\author{Manoel Horta Ribeiro}
\affiliation{
    \institution{EPFL}
    \country{Switzerland}
}
\email{manoel.hortaribeiro@epfl.ch}

\author{Shagun Jhaver}
\affiliation{
    \institution{Rutgers University}
    \country{USA}
}
\email{{shagun.jhaver@rutgers.edu}}

\author{Savvas Zannettou}
\affiliation{
    \institution{Max Planck Institute for Informatics}
    \country{Germany}
}
\email{szannett@mpi-inf.mpg.de}

\author{Jeremy Blackburn}
\affiliation{
    \institution{Binghamton University}
    \country{USA}
}
\email{jblackbu@binghamton.edu}

\author{Gianluca Stringhini}
\affiliation{
    \institution{Boston University}
    \country{USA}
}
\email{gian@bu.edu}

\author{Emiliano De Cristofaro}
\affiliation{
    \institution{University College London}
    \country{United Kingdom}
}
\email{e.decristofaro@ucl.ac.uk}

\author{Robert West}
\affiliation{
    \institution{EPFL}
    \country{Switzerland}
}
\email{robert.west@epfl.ch}

\renewcommand{\shortauthors}{Horta Ribeiro et al.}

\begin{abstract}
\textbf{When toxic online communities on mainstream platforms face moderation measures, such as bans, they may migrate to other platforms with laxer policies or set up their own dedicated websites.
Previous work suggests that \emph{within} mainstream platforms, community-level moderation is effective in mitigating the harm caused by the moderated communities.
It is, however, unclear whether these results also hold when considering the broader Web ecosystem.
Do toxic communities continue to grow in terms of their user base and activity on the new platforms? 
Do their members become more toxic and ideologically radicalized?
In this paper, we report the results of a large-scale observational study of how problematic online communities progress following community-level moderation measures.
We analyze data from r/The\_Donald and r/Incels, two communities that were banned from Reddit and subsequently migrated to their own standalone websites.
Our results suggest that, in both cases, moderation measures significantly decreased posting activity on the new platform, reducing the number of posts, active users, and newcomers. 
In spite of that, users in one of the studied communities (r/The\_Donald) showed increases in signals associated with toxicity and radicalization, which justifies concerns that the reduction in activity may come at the expense of a more toxic and radical community.
Overall, our results paint a nuanced portrait of the consequences of community-level moderation and can inform their design and deployment.}
\end{abstract}


\maketitle

\begin{center}
\large
\textcolor{red}{\textbf{This paper has been accepted at CSCW 2021, please cite accordingly.}
}\end{center}

\newpage

\input{body}

\bibliographystyle{acm}
\bibliography{refs}

\end{document}
\endinput

%% file: dlabmacros.tex
\usepackage[utf8]{inputenc}
\usepackage[T1]{fontenc}
\usepackage{hyphenat}
\usepackage{xspace}
\usepackage{amsmath}
\usepackage{amsfonts}
\usepackage{hyperref}
\usepackage{url}
\usepackage{booktabs}
\usepackage{multirow}
\usepackage{makecell}
\usepackage{caption}
\usepackage{minibox}
\usepackage{bbm}
\usepackage{graphicx}
\usepackage{balance}
\usepackage{mathtools}
\usepackage{color}
\usepackage{marvosym}
\usepackage{ifthen}
\usepackage{textcomp}
\usepackage{enumitem}
\usepackage{verbatim}
\usepackage{algorithm}
\usepackage{algorithmic}
\usepackage{numprint}
\usepackage{balance}

\usepackage{amsthm}
\theoremstyle{plain}

\newcommand{\chatoDisplayMode}[1]{#1}

\definecolor{MyRed}{rgb}{0.6,0.0,0.0} 
\definecolor{MyBlack}{rgb}{0.1,0.1,0.1} 
\newcommand{\inred}[1]{{\color{MyRed}\sf\textbf{\textsc{#1}}}}
\newcommand{\frameit}[2]{
  \begin{center}
  {\color{MyRed}
  \framebox[.9\columnwidth][l]{
    \begin{minipage}{.85\columnwidth}
    \inred{#1}: {\sf\color{MyBlack}#2}
    \end{minipage}
  }\\
  }
  \end{center}
}

\newcommand{\note}[2][]{\chatoDisplayMode{\def\@tmpsig{#1}\frameit{{\Pointinghand} Note}{#2\ifx \@tmpsig \@empty \else \mbox{ --\em #1}\fi}}}
\newcommand{\todo}[2][]{\chatoDisplayMode{\def\@tmpsig{#1}\frameit{{\Writinghand} To-do}{#2\ifx \@tmpsig \@empty \else \mbox{ --\em #1}\fi}}}

\newcommand{\abbrevStyle}[1]{#1}

\newcommand{\ie}{\abbrevStyle{i.e.}\xspace}
\newcommand{\eg}{\abbrevStyle{e.g.}\xspace}
\newcommand{\cf}{\abbrevStyle{cf.}\xspace}

\newcommand{\vs}{\abbrevStyle{vs.}\xspace}
\newcommand{\etc}{\abbrevStyle{etc.}\xspace}

\newcommand{\Secref}[1]{Sec.~\ref{#1}}

\newcommand{\Tabref}[1]{Table~\ref{#1}}
\newcommand{\Figref}[1]{Fig.~\ref{#1}}

\newcommand{\xhdr}[1]{\vspace{1.7mm}\noindent{{\bf #1.}}}

\newcommand{\textcite}[1]{\citeauthor{#1} \shortcite{#1}}

\newcommand{\hide}[1]{}

\makeatletter
\newcommand{\iffont}[2]{\ifthenelse{\equal{\f@family}{#1}}{#2}{}}
\makeatother

\iffont{ptm}{
  \usepackage{mathptmx}

  \DeclareSymbolFont{greek}{OML}{cmm}{m}{n}
  \DeclareMathSymbol{\alpha}{\mathalpha}{greek}{"0B}
  \DeclareMathSymbol{\beta}{\mathalpha}{greek}{"0C}
  \DeclareMathSymbol{\gamma}{\mathalpha}{greek}{"0D}
  \DeclareMathSymbol{\delta}{\mathalpha}{greek}{"0E}
  \DeclareMathSymbol{\epsilon}{\mathalpha}{greek}{"0F}
  \DeclareMathSymbol{\zeta}{\mathalpha}{greek}{"10}
  \DeclareMathSymbol{\eta}{\mathalpha}{greek}{"11}
  \DeclareMathSymbol{\theta}{\mathalpha}{greek}{"12}
  \DeclareMathSymbol{\iota}{\mathalpha}{greek}{"13}
  \DeclareMathSymbol{\kappa}{\mathalpha}{greek}{"14}
  \DeclareMathSymbol{\lambda}{\mathalpha}{greek}{"15}
  \DeclareMathSymbol{\mu}{\mathalpha}{greek}{"16}
  \DeclareMathSymbol{\nu}{\mathalpha}{greek}{"17}
  \DeclareMathSymbol{\xi}{\mathalpha}{greek}{"18}
  \DeclareMathSymbol{\pi}{\mathalpha}{greek}{"19}
  \DeclareMathSymbol{\rho}{\mathalpha}{greek}{"1A}
  \DeclareMathSymbol{\sigma}{\mathalpha}{greek}{"1B}
  \DeclareMathSymbol{\tau}{\mathalpha}{greek}{"1C}
  \DeclareMathSymbol{\upsilon}{\mathalpha}{greek}{"1D}
  \DeclareMathSymbol{\phi}{\mathalpha}{greek}{"1E}
  \DeclareMathSymbol{\chi}{\mathalpha}{greek}{"1F}
  \DeclareMathSymbol{\psi}{\mathalpha}{greek}{"20}
  \DeclareMathSymbol{\omega}{\mathalpha}{greek}{"21}
  \DeclareMathSymbol{\varepsilon}{\mathalpha}{greek}{"22}
  \DeclareMathSymbol{\vartheta}{\mathalpha}{greek}{"23}
  \DeclareMathSymbol{\varpi}{\mathalpha}{greek}{"24}
  \DeclareMathSymbol{\varrho}{\mathalpha}{greek}{"25}
  \DeclareMathSymbol{\varsigma}{\mathalpha}{greek}{"26}
  \DeclareMathSymbol{\varphi}{\mathalpha}{greek}{"27}
  \DeclareSymbolFont{otone}{OT1}{cmr}{m}{n}
  \DeclareMathSymbol{\Gamma}{\mathalpha}{otone}{0}
  \DeclareMathSymbol{\Delta}{\mathalpha}{otone}{1}
  \DeclareMathSymbol{\Theta}{\mathalpha}{otone}{2}
  \DeclareMathSymbol{\Lambda}{\mathalpha}{otone}{3}
  \DeclareMathSymbol{\Xi}{\mathalpha}{otone}{4}
  \DeclareMathSymbol{\Pi}{\mathalpha}{otone}{5}
  \DeclareMathSymbol{\Sigma}{\mathalpha}{otone}{6}
  \DeclareMathSymbol{\Upsilon}{\mathalpha}{otone}{7}
  \DeclareMathSymbol{\Phi}{\mathalpha}{otone}{8}
  \DeclareMathSymbol{\Psi}{\mathalpha}{otone}{9}
  \DeclareMathSymbol{\Omega}{\mathalpha}{otone}{10}
  \DeclareSymbolFont{syms}{OML}{cmm}{m}{it}
  \DeclareMathSymbol{\partial}{\mathord}{syms}{"40}
  \DeclareMathAlphabet{\mathbold}{OML}{cmm}{b}{it}
  \DeclareSymbolFont{largesymbols}{OMX}{cmex}{m}{n}

}

%% file: body.tex
\section{Introduction}

\noindent
\begin{center}
    \textbf{Warning: this work quotes slur terms that some may find offensive.}
\end{center}
\vspace{1mm}

The term ``content moderation'' is commonly associated with the process of screening the appropriateness of user-generated content, as well as imposing penalties on \emph{users} who break the rules~\cite{roberts2019behind}. 
However, social networking platforms sometimes host entire \emph{communities} that systematically defy regulations.
There, host platforms oftentimes ban or limit the functionalities of one or several online communities.
This has happened, for example, when Reddit decided that not all communities were welcome on the platform~\cite{dewey_these_2015} and banned subreddits like \texttt{r/FatPeopleHate} and \texttt{r/transfags}.
Also, more recently, following the 2021 storming of the United States Capitol, groups supporting far-right ideologies and the QAnon movement have been banned across different mainstream social media platforms~\cite{brewster_extremists_2020}.

The extent to which platforms should be the judges, juries, and executioners of these interventions is a topic of heated debate and has prompted experiments of governance models with societal participation~\cite{fb_news_welcoming_2020}.
There, platforms outsource some of their policy decisions---\eg, should we ban an online movement from our platform?---to a panel of experts (\eg, journalists, politicians, lawyers) representing the public interest~\cite{almeida2020digital}.

\begin{figure}[t]
 \centering
 \includegraphics[width=0.6\linewidth]{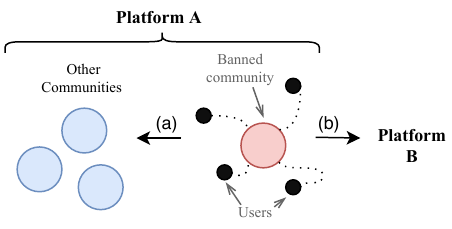}
 \caption{\textbf{Motivation:} As a result of community-level bans, users from affected communities may choose to \textit{(a)}~participate in other communities on the same platform or \textit{(b)}~migrate to an alternative, possibly fringe platform where their behavior is considered acceptable.
 Scenario \textit{a}, on which most prior work has focused, is more amenable to data-driven analysis. The present paper, on the contrary, focuses on the harder-to-analyze scenario \textit{b}.
 }
 \label{fig:motivation}
\end{figure}

Nonetheless, we are still left with answering a preceding question:
is community-level moderation effective to begin with?
We argue why this is not obvious, visually, in \Figref{fig:motivation}, depicting possible decisions (\emph{a} and \emph{b}) that users (the black dots) associated with a recently banned toxic%
\footnote{We use `toxic' as an umbrella term to refer to socially undesirable content: sexist, racist, homophobic, or transphobic posts, targeted harassment, and conspiracy theories that target racial or political groups.} 
community may take.
The users may
\emph{(a)}~continue to be active on the same platform and participate in other groups and communities there, or
\emph{(b)}~abandon the platform altogether and migrate to a different platform. 
In both scenarios, community-level moderation could have unintended consequences.
In scenario \emph{a}, the moderation measure could set loose an army of trolls across the platform, creating issues in other communities or \emph{new} problematic communities~\cite{chandrasekharan_you_2017}.
In scenario \emph{b}, the ban could unintentionally strengthen an alternative platform (\eg, 4chan or Gab) where problematic content goes largely unmoderated~\cite{mathew_temporal_2019}. 
From the new platform, the harms inflicted by the toxic community on society could be even higher.

Previous work has addressed the \textit{``within platform''} concern. 
Chandrasekharan et al.~\cite{chandrasekharan_you_2017} and Saleem and Ruths~\cite{saleem_aftermath_2018} studied what happened following Reddit's 2015 bans, finding that users who remained on the platform drastically decreased their usage of hate speech and that counter-actions taken by users from the banned subreddits were promptly neutralized.
More broadly, Rajadesingan et al.~\cite{rajadesingan_quick_2020} showed that, when ``toxic users'' migrate to healthy communities, they reduce their toxicity levels.

Nevertheless, the concern that migrations to an \emph{alternative} platform would strengthen the toxic communities or make them more ideologically radical is still largely unexplored. 
Existing work suggests that, in the wake of community-level moderation, users actively seek out, and migrate to, alternative websites where their speech will not be censored~\cite{shen_discourse_2019,newell_user_2016}.
However, partly due to the data collection challenges posed by cross-platform studies, quantitative work on the consequences of community-level moderation \emph{across platforms} has remained at the simulation level~\cite{johnson_hidden_2019}.

\xhdr{Present work}
This paper presents an observational
study of the efficacy of community-level moderation across platforms.
We examine two popular communities that were originally created and grew on Reddit, r/The\_Donald and r/Incels.
Faced with sanctions from the platform,
they created their own standalone websites---thedonald.win and incels.co---and encouraged their Reddit user base to mass migrate to the new websites.
To assess whether community\hyp{}level moderation measures were effective in reducing the negative impact of these communities  (which we refer to as \emph{TD} and \emph{Incels}, respectively), we study how they progressed following their platform migrations. 
More specifically, we ask:

\begin{itemize}
 \item[\textbf{RQ1}] Have the communities retained their activity levels and their capacity to attract new members following the migration to a new platform?
 \item[\textbf{RQ2}] Have the communities become more toxic or ideologically radical following the migration to a new platform?
\end{itemize}

\noindent
Both dimensions are crucial to assess whether community\hyp level moderation measures were truly effective.
If the communities simply ``changed addresses'' and grew larger and more toxic on the new platforms, the moderation measures may have actually increased their capacity to harm society as well as their own members; \eg, outside of Reddit, these communities might orchestrate online harassment campaigns more effectively or disseminate more hate speech.

\xhdr{Materials and methods} 
To study how migrations affect communities, we leverage over 6 million posts made by more than 138 thousand users pooled across the platforms before (Reddit) and after (standalone websites) the migration event.
We extract activity-related signals, such as the number of posts, active users, and newcomers, as well as content-related signals, such as algorithmically derived ``toxicity scores,'' that aim to identify behaviors indicative of user radicalization, such as fixation and group identification~\cite{cohen_detecting_2014}.
Employing quasi-experimental setups, including matching and regression discontinuity analysis, we study these signals from a \textit{community-level perspective}, analyzing how daily activity and overall content changed, and from a \textit{user-level perspective}, examining how the behavior of individual users changed following platform migrations. 

\xhdr{Summary of findings}
Analyzing activity levels and the inflow of newcomers to the communities (\textbf{RQ1}), we find that the moderation measures significantly reduced the overall number of active users, newcomers, and posts in the new communities compared to the original ones. 
However, individually, users posted more often on the alternative platforms.
A closer look at the users whom we managed to match before \vs after the migration suggests that this increase in \textit{relative activity} is due more to self-selection rather than behavior change. Users who migrated were more active in the original platform, and their activity dropped on a user level.

Analyzing changes in the content being posted in the communities following the migration (\textbf{RQ2}), we find evidence that users in the TD community became more toxic, negative, and hostile when talking about ``objects of fixation'' (\eg, democrats, leftists).
Changes in the usage of third-person plural (\eg, ``they'') and first-person plural (\eg, ``we'') pronouns also indicate an increase in ingroup identification and in othering language.
For the Incel community, we find that changes tend to be statistically non-significant.

\xhdr{Implications}
Our analysis suggests that community-level moderation measures decrease the capacity of toxic communities to retain their activity levels and attract new members, but that this may come at the expense of making these communities more toxic and ideologically radical.
Therefore, as platforms moderate, they should consider their impact not only on their own websites and services, but in the context of the Web as a whole.
Toxic communities respect no platform boundary, and thus, platforms should consider being more proactive in identifying and sanctioning toxic communities before they have the critical mass to migrate to a standalone website.
Overall, we expect that our nuanced analysis will aid stakeholders to take moderation decisions and make moderation policies in an evidence-based fashion.

\section{Background and related work}

\subsection{Community-level moderation on Reddit}
Reddit employs two community-wide moderation measures: \emph{quarantining} and \emph{banning}. 
When a community is quarantined, it stops appearing in Reddit's search results and front page. 
Moreover, users who attempt to access quarantined subreddits (directly through their URLs) are met with a splash page warning them of the shocking or offensive content contained inside.
In contrast, banning a community makes it inaccessible and removes all its prior posts.
Quarantining frequently precedes banning, so in practice, it serves as a warning for the subreddit to reform itself.

The history of community-level moderation in Reddit dates back to 2015 when Reddit banned five subreddits for infringing their anti-harassment policy~\cite{dewey_these_2015}.
Newell et al.~\cite{newell_user_2016} studied how these bans led users to migrate towards alternative platforms (\eg, Voat).
Using a mix of self-reported statements and large-scale data analysis, they identified reasons why users left Reddit and found that alternative platforms struggled to attain the same diversity of communities as Reddit. 
The effects of these bans \emph{within} Reddit were also extensively studied~\cite{saleem_aftermath_2018,chandrasekharan_you_2017}, as previously discussed.
Overall, findings from these studies suggest that the bans worked for Reddit:
they led to sustained reduced interaction of users with the Reddit platform; 
users who stayed became less toxic after they migrated to other communities within Reddit; 
and counter-actions taken by users (\eg, creating alternative subreddits) were not effective.

\subsection{Communities of interest} 

\xhdr{TD} The r/The\_Donald subreddit (TD) was created on 27 June 2015 to support the then presidential candidate Donald Trump in his bid for the 2016 U.S.\ Presidential election. 
The discussion board, linked with the rise of the Alt-right movement at large, has been denounced as racist, sexist, and islamophobic~\cite{lyons2017ctrl}.
Its members often engaged in ``political trolling,'' harassing Trump's opponents, promoting satirical hashtags, and creating memes with pro-Trump and anti-Clinton propaganda~\cite{flores-saviaga_mobilizing_2018}.
TD is also known for spreading unsubstantiated conspiracy theories like Pizzagate~\cite{ohlheiser_2016} and the Seth Rich murder conspiracy~\cite{gilmour_2017}.

\begin{figure}[t]
 \centering
 \includegraphics[width=0.9\linewidth]{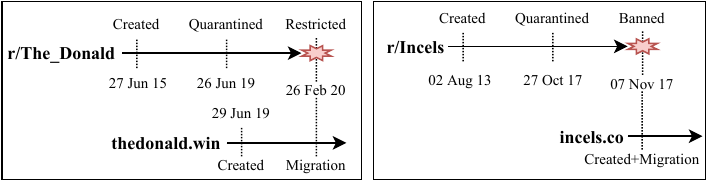}
 \caption{\textbf{Timelines}: We depict the dates of creation, quarantining, and banning for the two communities studied here.}
 \label{fig:subreddits_timelines}
\end{figure}

We depict important events in TD's history in \Figref{fig:subreddits_timelines}.
The subreddit was quarantined in mid-June 2019 for violent comments, and on 26 February 2020, Reddit administrators removed a number of TD's moderators, and the community was placed under a ``restricted mode,'' which removed the ability of most of its users to post.
Months after the subreddit became inactive, it was banned in late June 2020.
While these moderation measures were taking place, TD users were actively organizing a ``plan B.''
In 2017, its members were already considering migrating to alternative platforms~\cite{reddit_post_td}, and in 2019, after getting quarantined, moderators created a backup site, thedonald.win, that was promoted in the subreddit using stickied posts~\cite{thedonaldpost0} (\ie, always shown among the first in the feed for the community).
TD users continued using the subreddit until the community became ``restricted.'' 
Then, they largely flocked to the alternative website~\cite{timberg_reddit_2020}.
Note that, although TD was eventually banned, we focus here on its ``restriction,'' since it was this measure that halted user participation and ignited the community migration.
 
\xhdr{Incels}
The r/Incels subreddit was created in August 2013. 
Short for involuntary celibates, it was a community built around ``The Black Pill,'' the idea that looks play a disproportionate role in finding a relationship and that men who do not conform to beauty standards are doomed to rejection and loneliness~\cite{lilly2016world,ging2017alphas}.
Incels rose to the mainstream due to their association with mass murderers~\cite{hoffman2020assessing} and their obsession with plastic surgery~\cite{hines_how_2019}.
The community has been linked to a broader set of movements~\cite{lilly2016world,ribeiro_evolution_2020} referred to as the ``Manosphere,'' which espouses anti-feminist ideals and sees a ``crisis in masculinity.''
In this world view,  men and not women are systematically oppressed by modern society.
Lately, specialists have also suggested that these communities may play an important role in radicalizing disenfranchised men and producing ideological echo chambers that promote violent rhetoric~\cite{hoffman2020assessing}.

The r/Incels subreddit grew swiftly in early 2017, reaching over 3,000 daily posts~\cite{ribeiro_evolution_2020}. 
Shortly after, in late October 2017, it was quarantined and then banned two weeks later~\cite{solon_incel_2017}.
In an interview for a podcast~\cite{kates_statusmaxxing_2019}, one of the subreddits' former core members, \textit{seargentincel}, mentions that he had already discussed moving the community outside of Reddit with moderators.
According to him, when the subreddit was banned, he created the standalone website incels.co, and former r/Incels members quickly organized the migration in Discord channels.
Again, we provide exact dates for relevant events in \Figref{fig:subreddits_timelines}.

\xhdr{Choice of communities} 
We study these two communities for two main reasons.
First, due to their importance: they have a large number of members and have impacted society at large, \eg, w.r.t.~conspiracy theories~\cite{gilmour_2017} and real-world violence~\cite{hoffman2020assessing}.
Second, these are communities whose migrations were backed by community leaders, and that migrated to other public websites. 
Had the members of these communities spread to a loosely connected network of private channels (\eg, on Telegram), there would be several additional technical and ethical research challenges.

\subsection{Toxicity and radicalization online}

Internet platforms experience a myriad of toxic behaviors such as incivility~\cite{borah2014does}, harassment \cite{blackwell2017classification,jhaver2018blocklists}, trolling~\cite{cheng2015antisocial} and cyberbullying~\cite{kwak2015exploring}. 
In recent years, researchers have explored the dynamics of such behaviors online aided by automatic methods~\cite{mathew_temporal_2019, ribeiro_evolution_2020}. 
Broadly, the methods employed fall under one of two categories. 
They either \emph{(a)} count hate-related or toxicity-related words (\eg, using HateBase~\cite{hatebase_hatebase_2018}); or \emph{(b)} deploy machine-learning based methods to classify comments as toxic or as hateful (\eg, Google's Perspective API~\cite{jigsaw2018perspective}).
Methods differ in what they intend to measure: some aim to measure ``hate speech,'' while others ``toxicity.''
While these concepts differ tremendously, research has suggested that measuring hate speech through text is difficult due to its contextual nature, and that machine learning classifiers struggle to distinguish between offensive and hateful speech~\cite{horta_ribeiro_characterizing_2018, davidson2017automated}.

Intertwined with online toxicity are movements and ideologies that engage in harassment campaigns and real-world violence, as well as espouse hateful views towards minorities~\cite{lewis2018alternative,massanari2017gamergate}. 
Social networks have been identified as places where individuals are exposed and eventually adhere to such fringe movements~\cite{ribeiro2020auditing}.
In this direction, the work of Grover and Mark~\cite{grover_detecting_2019}, also on Reddit, is particularly relevant, as their work suggests that behaviors indicative of radicalization such as fixation and group identification may be captured through automated text analysis.
We extend their methodology to assess the changes in user-generated \textit{content} following the migrations, using the same word categories (derived from Linguistic Inquiry and Word Count, or LIWC~\cite{pennebaker2007computerized}) and developing, for the Incel and TD communities, custom-built ``fixation dictionaries'' that contain terms serving as objects of fixation in the communities (\eg \emph{leftist} for TD, \emph{feminism} for Incels).
Additionally, we use the Perspective API to measure how toxicity in these communities changed post-migration.

The suitability of using models from the Perspective API as toxicity sensors has been explored in previous work.
Rajadesingan et al.~\cite{rajadesingan_quick_2020} found that, for Reddit political communities, the performance of the classifier is similar to that of a human annotator, while Zannettou et al.~\cite{zannettou2020measuring} found that Perspective's ``Severe Toxicity'' model outperforms alternatives like HateSonar~\cite{davidson2017automated}.
Perspective has been shown to be biased against comments mentioning marginalized subgroups and for comments posted in African\- American English~\cite{sap2019risk}.
We find no compelling reason to believe that these biases may impact the post-migration changes in the toxicity of the communities studied.

Lastly, it is worth stressing that the utility of understanding toxicity in online communities goes beyond the study of fringe or troublesome communities.
In the context of peer production communities, Carillo and Marsany~\cite{carillo2016dose} have discussed toxicity drawing notions from ecology and toxicology: exposure to ``toxic'' content without the appropriate ``defense mechanisms'' would harm the productivity of online communities.
In this paradigm, efforts to understand, pro-actively detect, and quickly act against antisocial or toxic behavior would be key in maintaining healthy online communities, directions that have been empirically explored by previous work~\cite{cheng2017anyone,raman2020stress,zhang2018conversations}.

\subsection{Relation with prior work}

Overall, previous research discussed above has examined the efficacy of community-level moderation \emph{within} Reddit~\cite{chandrasekharan_you_2017,saleem_aftermath_2018} and analyzed cross-platform migrations that ensued~\cite{newell_user_2016}. 
Our work takes a significant step further, by assessing the efficacy of these interventions in a new direction.
Given that communities \emph{do} migrate following moderation measures, we study if these measures are effective when considering the development of the communities outside of their original platform.
To do so, we draw from a rich literature of existing work on online toxicity and its relationship with behaviors indicative of radicalization~\cite{grover_detecting_2019}, as well as on previous studies analyzing the communities at hand~\cite{flores-saviaga_mobilizing_2018, ribeiro_evolution_2020}.

\section{Materials and methods}

\subsection{Data collection}
We collect data from both Reddit (for the period \emph{before} migrations) and standalone websites (for the period \emph{after}).

\xhdr{Reddit} 
To collect Reddit data, we use Pushshift~\cite{baumgartner_pushshift_2020}, a service that performs large-scale Reddit crawls. 
We collect all submissions and comments made on r/The\_Donald and r/Incels, starting from 120 days before the moderation measure, and until its date.
Specifically, for r/Incels, we collect data between 10 July 2017 and 7 November 2017; for r/The\_Donald, between 29 October 2019 and 26 February 2020.
Overall, we collect around 3 million comments in 260K submissions (or ``threads'') from both subreddits (see Table~\ref{tab:datasets}). 

\begin{table}[t]
\small
\centering
\caption{Overview of our datasets. }
\input{images/overview_datasets}
\label{tab:datasets}
\end{table}

\xhdr{Standalone websites}
We additionally implement and use custom Web crawlers to collect data from the standalone websites (incels.co and thedonald.win).
For each, we collect all submissions and comments posted for a period of 120 days after the community-level moderation measure.
Specifically, for incels.co, we collect data between 7 November 2017 and 6 March 2018; for thedonald.win, between 26 February 2020 and 24 June 2020.
Overall, we collect over 2.5 million comments and submissions from thedonald.win and over 400K comments and submissions from incels.co. 
In the rest of the paper, to ease presentation, we refer to both submissions and comments as ``posts.''

\subsection{User analysis}
\label{sec:methods-user}
We briefly describe our methods for matching users across platforms and for analyzing newcomers.

\xhdr{Matched Users}
To better understand changes at the user-level, we also carry out analyses with matched users, finding pairs of users with the exact same username on both Reddit and the standalone websites.
We consider that these users are the same individuals in the two platforms, an assumption backed by anecdotal evidence from within the communities (thedonald.win even had a feature to reserve your Reddit username~\cite{thedonaldpost1}) and by previous research~\cite{newell_user_2016}.
Allowing for upper/lower-case differences, using this method, we were able to match 8,651 users between r/The\_Donald and thedonald.win (around 20\% of the user base of the latter) and 286 users between r/Incels and incels.co (around 13\%).

\xhdr{Newcomers}
We estimate the inflow of newcomers in each community considering both the pre- and post-migration period by counting the daily number of posts with usernames never before observed. For instance, if a user \textit{X} posted in thedonald.win on the 1st of March of 2020, and no user with such username posted before in either thedonald.win or r/The\_Donald, we would consider him or her a newcomer.
Note that here, if we used only the data from 120 days before and after the migration, we would observe a spike in newcomers at the beginning of the study period. To prevent that, for each community, we additionally download all history available in Pushshift to act as a buffer.

\subsection{Content analysis}
\label{sec:methods-content}

To understand the impact of platform migration on the content being produced by the communities, we use text-based signals  associated with toxicity and user radicalization~\cite{grover_detecting_2019,mathew_temporal_2019}.

\xhdr{Fixation dictionary} 
We generate a \emph{fixation dictionary} for each of the communities, selecting terms related to their ``objects of fixation.'' 
More specifically, we:
(1)~select terms that are more likely to occur in the communities of interest as compared to Reddit in general, and
(2)~manually curate these terms, selecting those that are related to these communities' objects of fixation (\eg, \textit{women} and \textit{feminism} for Incels). 
To obtain the list of terms, we extract words from the communities of interest and from a 1\% random sample of Reddit for a period of one month (immediately prior to the study period previously described). 
We exclude bot-related messages (\eg, auto-moderation), stop-words, and words that occurred fewer than 50 times,
and calculate the log-ratio between the frequency of a keyword in the communities being studied and on Reddit in general.
From this, we obtain, for each community, the 250 terms that have the highest relative occurrence.
Then, to build the fixation dictionary, three authors of this paper (all familiar with the communities at hand) discussed each term and came to an agreement on whether or not that term was an object of fixation.
Table~\ref{tab:fixation} reports the terms in our fixation dictionary for each community; be advised that the terminology in this table is offensive. 

\begin{table}[t]
 \centering
 \small
 \caption{Fixation dictionaries.}
 \vspace{-2mm}
 \label{tab:fixation}
 \input{images/fix_dict}
\vspace{-3mm}
\end{table}

\xhdr{Toxicity score} 
To analyze content toxicity, we use Google's Perspective API~\cite{jigsaw2018perspective}, an API consisting of machine learning models trained on manually annotated corpora of text.
More specifically, we employ the ``Severe Toxicity'' model which allows us to assess how likely (on a scale between 0 and 1) a post is to be ``rude, disrespectful, or unreasonable and is likely to make you leave a discussion.'' 
This model is also trained specifically to not classify benign usage of foul language as toxic.

\xhdr{LIWC}
We measure changes in word choice using the Linguistic Inquiry and Word Count (LIWC) tool~\cite{tausczik2010psychological}.
LIWC consists of various dictionaries (in total 4.5K words) to classify words into over 70 categories, including general characteristics of posts (\eg, word count), linguistic components (\eg, adverbs), psychological processes (\eg, cognitive processes), and non-psychological processes (\eg, pronouns).
In this work, we study changes for the following (aggregated) LIWC 2015 categories:
(1)~Negative Emotions: sum of the \emph{Anger}, \emph{Anxiety}, and \emph{Sadness} LIWC categories;
(2)~Hostility: sum of \emph{Anger}, \emph{Swear}, and \emph{Sexual} LIWC categories.
(3)~Pronouns: we focus on the usage of third-person plural (\eg, ``they''), and first person plural (\eg, ``we'') pronouns.

\xhdr{Mapping signals to warning behaviors} 
These different signals, as well as their combinations, have been described as warning behaviors of ideological radicalization.
We focus on two warning behaviors described by Cohen et al.~\cite{cohen_detecting_2014}:
(1)~\textit{Fixation:}
a pathological preoccupation with a person or cause that is increasingly expressed with negative and angry undertones; and
(2)~\textit{Group Identification:} strong identification and moral commitment to the ingroup and distancing from the outgroup.
To study changes in \textit{fixation}, we analyze our fixation dictionaries along with \emph{Toxicity scores} and the word categories \textit{Negative Emotions} and \textit{Hostility}.
To study changes in \textit{group identification}, we study changes in the usage of pronouns as measured by LIWC.
These choices were motivated, as discussed earlier, by previous work by Grover and Mark~\cite{grover_detecting_2019}.

\subsection{Ethics and reproducibility} 
In this work, we only used data publicly posted on the Web and did not
(1)~interact with online users in any way, nor
(2)~simulate any logged-in activity on Reddit or the other platforms.
When we matched users on Reddit and the fringe platforms, we did not attempt to gain any information about users' personal identities.
Anonymized reproducibility data and code are available at \url{https://doi.org/10.5281/zenodo.5171068}
We stress that the data is provided without the usernames or the actual text posted (i.e., only the signals extracted). 
We believe that this makes de-anonymization harder than crawling the standalone websites and downloading existing Reddit dumps. 
These steps follow previous work studying toxic communities on Reddit~\cite{rajadesingan_quick_2020}, and we believe they minimize the potential harms associated while ensuring the study is reproducible. 
Additionally, we note that we only do \emph{exact matching} on publicly available data, while not singling out any individual user, and thus we believe we are not infringing on reasonable privacy expectations. 

\section{Changes in activity levels}
\label{sec:activity}

In this section, we measure how the community-level moderation measures changed posting activity levels and the capacity of the two communities to attract newcomers (\textbf{RQ1}). 
We do so from two different perspectives. 
First, we aggregate our data on a daily basis, inspecting \textit{community-level} changes in the number of posts, active users and newcomers. 
Next, we zoom in to the \textit{user-level} and examine how individual users' behavior changed post-migration.

\subsection{Community-level trends}

\begin{figure*}
 \centering
\includegraphics[width=\textwidth]{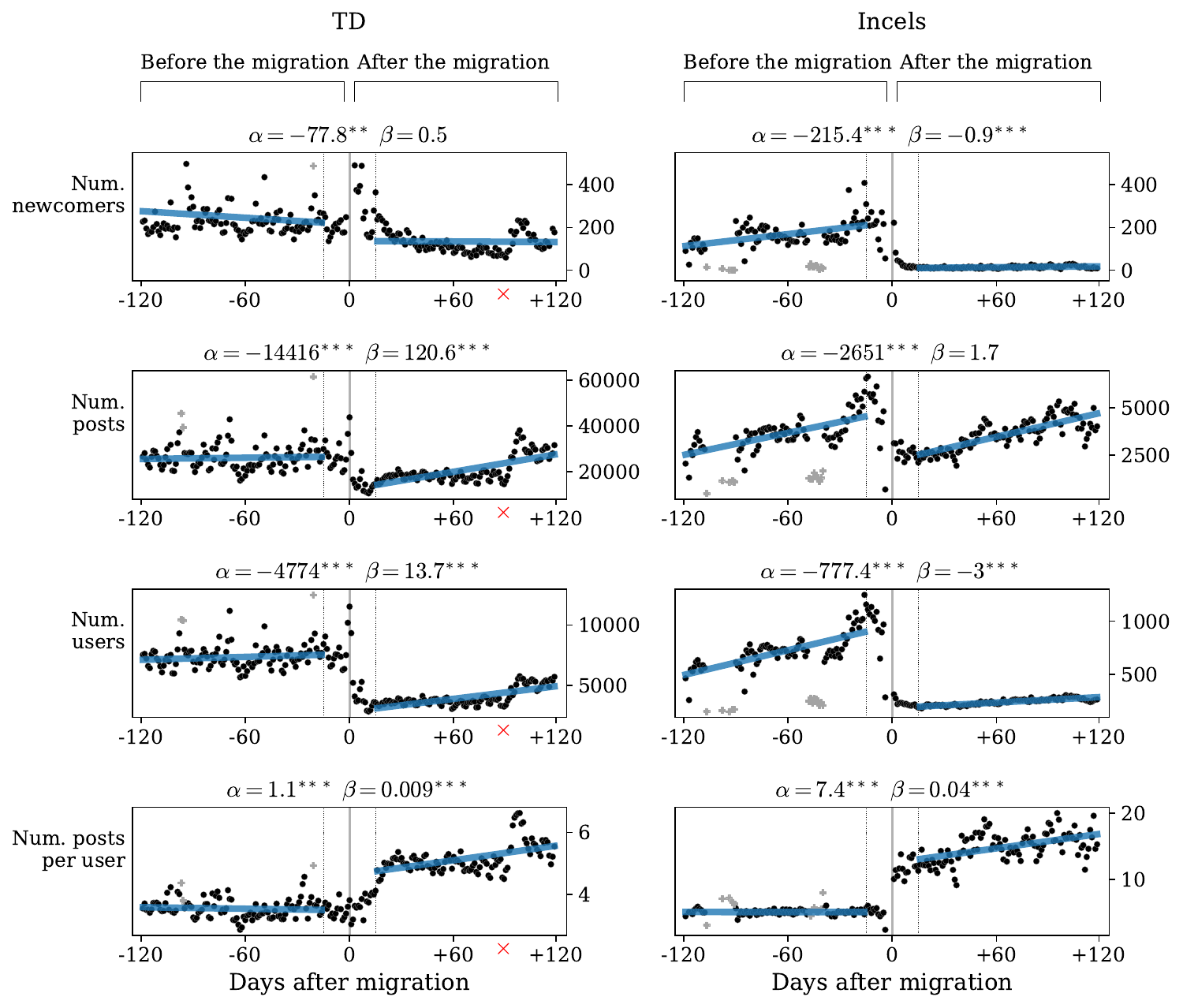}
\caption{
\textbf{Activity levels:}
Daily activity statistics for the TD community (left) and the Incel community (right) 120 days before and after migrations. 
Dots represent the daily average for each statistic, and the blue lines depict the model fitted in the regression discontinuity analysis.
The migration date and a grace period around it (used in the model) are depicted as solid and dashed gray lines, respectively.
Gray crosses represent days where the Pushshift ingest had issues, or where there was a large volume of spam-like content.
On top of each subplot, we report the coefficients associated with the moderation measure in the model ($\alpha$ and $\beta$). Coefficients for which $p<$ 0.001, 0.01, and 0.05 are marked with ***, **, and *, respectively.
For the TD community, we mark the killing of George Floyd (on 25 May 2020), with a red cross ($\color{red} \times$) close to the x-axis.
}
\label{fig:trends}
\end{figure*}

\Figref{fig:trends} shows the daily number of newcomers, posts, and active users in each community before and after the migrations for both the TD and the Incel community. 
Note that we consider data from both the subreddit and the fringe platform users migrated towards. 

To gain a better understanding of the overall trends, we perform a regression discontinuity analysis for each statistic in each community.
We employ a linear model:
\begin{equation}
\label{eq:linear}
 y_t =  \alpha_0 + \beta_0 t  + \alpha i_t + \beta i_t t,
\end{equation}
\noindent
where
$t$ is the date, which takes values between $-120$ and $+120$ and equals 0 in the day of the moderation measure;
$y_t$ is statistic we are modeling; 
and $i_t$ is an indicator variable equal to 1 for days following the moderation measure (\ie, $t>0$), and 0 otherwise.
Our model assumes that daily activity levels (for the different metrics) can be approximated by a line (defined by coefficients $\alpha_0$ and $\beta_0$), which, post-migration, can change both its intercept ($\alpha$) and its slope ($\beta$).
We analyze these changes to understand the impact of platform migrations on the communities at hand.

We exclude data from a ``grace period'' of 15 days before and after the moderation measure.%
\footnote{We stress that our results are robust to changes in this parameter: we have experimented with different window sizes (e.g. 7 and 21 days), obtaining largely the same results}
This accounts for the bursty behavior happening on user activity metrics in the days around the migration. 
For example, for newcomers, many of the users who migrated to the new website (thedonald.win or incels.co) choose new usernames, which creates a spike in the metric.
However, this initial spike is not interesting to capture the overall trend of newcomers in the website, and the grace period addresses that.
Additionally, there were a few days on which the Pushshift ingest had problems or where there was a large volume of spam-like content. 
The values for the statistics on these dates are depicted as gray crosses in \Figref{fig:trends} and were not considered to fit the models.
Both coefficients and 95\% CI for each parameter in the regressions are shown in \Tabref{tab:rdd}. 

\xhdr{Newcomers} 
The first row of \Figref{fig:trends} shows the number of daily newcomers in each community (as described in \Secref{sec:methods-user}). 
We find that, for both communities, there was a significant \emph{decrease} in the influx of newcomers following the migration.
The TD community saw a significant decrease of around 78 daily newcomers ($\alpha = -77.8$).
This represents a percent change of around $-30\%$ of the \emph{M}ean \emph{V}alue \emph{B}efore the community-level \emph{I}ntervention (referred to as \emph{MVBI} henceforth), \ie, the drop represents roughly 30\% of the average daily value in the pre-migration period. 
The decrease was even more substantial for the Incel community, which experienced around $215$ fewer newcomers a day ($\alpha = -215.4$),
roughly $-150\%$ of the \emph{MVBI} (note that the drop was, therefore, \emph{bigger} than the pre-migration average).
Furthermore, the Incel community had a significant increasing trend before the migration ($\beta_0 = 0.9$),
which was weakened in the post-migration period ($\beta = -0.9$).

\xhdr{Posts and users} 
The second and third rows of \Figref{fig:trends} show that both the total number of daily posts and daily posting users dropped significantly post-migration.
TD experienced a decrease of around $14.4$k daily posts
($\alpha=-14416$, $-55\%$ of the \emph{MVBI})
and of around 4.7k daily active users
($\alpha=-4774$, $-65\%$ of the \emph{MVBI}). 
In both cases, the slope became steeper after the migration, with a significant increase of around $121$ new posts a day ($\beta = 120.6$), and around $14$ additional active users a day ($\beta = 13.7$).
A possible explanation for this increase is that the killing of George Floyd (25 May 2020) and the demonstrations that ensued may have boosted participation on the platform, since the date coincides with a sharp rise in both statistics.
Repeating the regression analysis excluding the period after 24 May 2020, we find non-significant \textit{decreases} in the slope ($\beta$) for both statistics, which strengthens this hypothesis. 
We further discuss this confounder in \Secref{sec:discussion}.

For the Incel community, there were significant decreases of around $2.6$k posts a day 
($\alpha = -2651$, $-73\%$ of the \emph{MVBI}),
 and of around $777$ daily active users
($\alpha = -777.4$, $-116\%$ of the \emph{MVBI}). 
Looking at the trends for the number of active users, we find a significant positive trend across the whole period ($\beta_0=3.9$, see \Tabref{tab:rdd}) but the slope decreases significantly after the migration ($\beta = -3$).

\xhdr{Posts per user} 
The fourth row of \Figref{fig:trends} shows the daily average of the posts per user ratio.
Here, we find that the moderation measure \textit{significantly increased} relative activity. 
The TD community showed an increase in the number of daily posts per user of around $1$ extra posts per user ($\alpha=1.1$, $31\%$ of the \textit{MVBI});
for Incels, the increase was of around 7 extra posts per user 
($\alpha=7.4$, $123\%$ of the \textit{MVBI}).
In both cases there is also a significant increase in the trend ($\beta = 0.009$ for TD, and $\beta = 0.04$ for Incel).
This adds nuance to the overall scenario:  
although the activity in the communities is reduced, \textit{relative} to the number of users, it increases.

\subsection{User-level trends}

The analyses done so far paint a comprehensive picture of the changes in activity due to migration at the community-level. 
Yet, they do not disentangle the effects happening at the user-level.
We found that \emph{relative} activity increased (\ie, fewer users posted more often), but the underlying mechanism for this change is still unclear.
Users' activity may have indeed increased \emph{after} the migration, \ie, individually, each user who migrated might post more often on the fringe website, but the increase could also be due to self-selection: users who migrated following the moderation measure might have been more active to begin with. 

Understanding the reason behind this (relative) activity increase is important to evaluate the efficacy of the moderation measure. 
If the increase occurred because users became more active, the subset of users ``ignited'' by the moderation measure could cause even greater harm in the new platform. 
However, if the increase was only due to self-selection, we might consider the measure successful in decreasing the activity and reach of the communities.
To better understand the mechanism behind this activity increase, we perform an additional set of analyses inspecting what changed at the \emph{user level} post-migration.
To do so, we analyze the set of users before and after the migration, and additionally, the set of \textit{matched users} described in \Secref{sec:methods-user}. 

\begin{figure}[t]
\centering
\includegraphics[width=\linewidth]{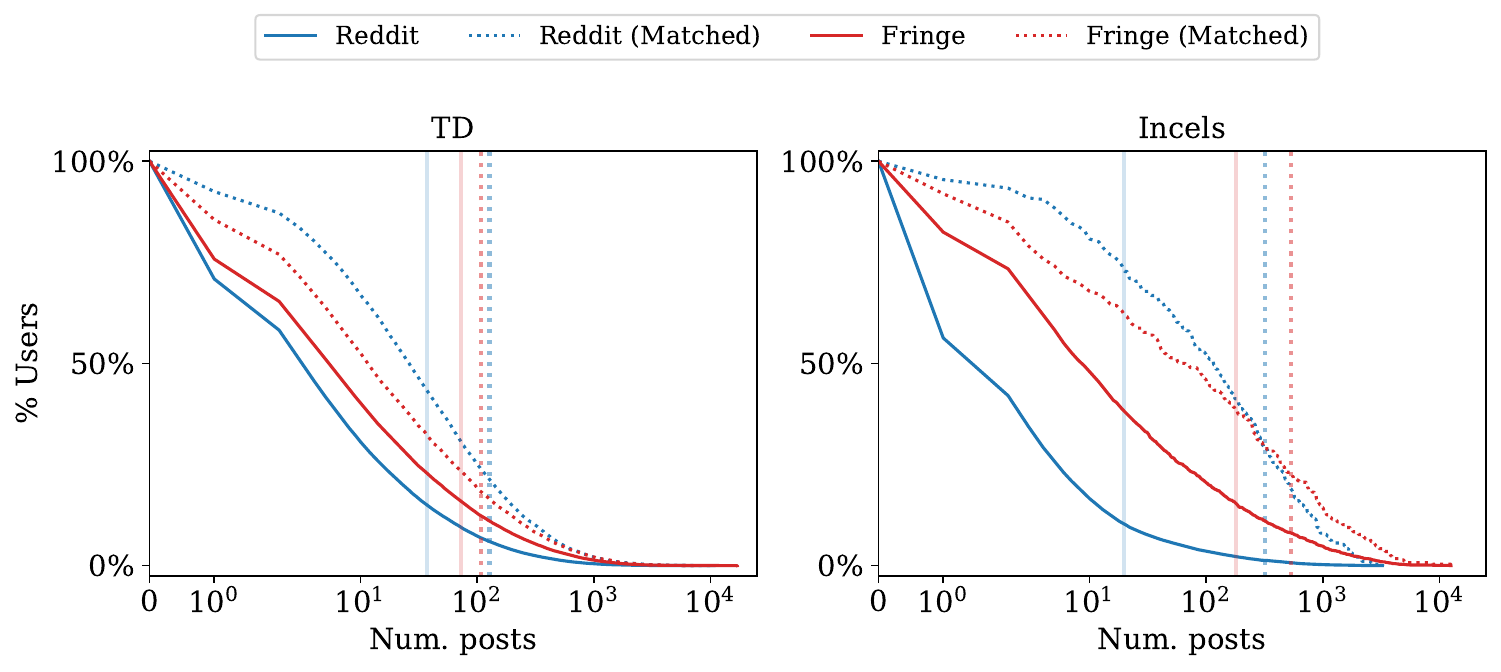}
\caption{
\textbf{CCDFs of posts per user:}
For each community, we depict the complementary cumulative distribution function (CCDF) of the number of posts per user for: 
(1) all users who posted in the 120 days \emph{before} (solid blue) and \emph{after} the moderation measure (solid red), and 
(2) users we managed to match based on username while they were still on Reddit (dashed blue) and on the fringe platform (dashed red).
The plot also depicts the mean value for each one of these populations as vertical lines in the same color/style scheme. 
Recall that the CCDF maps every value in the $x$-axis to the percentage of values in a sample that are bigger than $x$ (in the $y$-axis).
}
\label{fig:matched_summary}
\end{figure}

\xhdr{Comparing posts-per-user distributions} 
We begin by comparing the distribution of posts from matched users and the general population of users both before and after the migration.
\Figref{fig:matched_summary} depicts the complementary cumulative distribution function (CCDF) of the number of posts for all users (solid line) and for matched users (dashed line) both in the fringe communities (in red) and on Reddit (in blue).

Considering all users (solid lines), the CCDFs confirm our previous analysis, showing that users are more active in the fringe websites, since the red solid line is consistently above the blue solid line.
This is also captured by the mean of user activity in fringe communities, which is of 
73 posts per user (95\% CI: [70, 76.3]\footnote{Confidence intervals were calculated through bootstrapping.}) for the TD community, and of 
180.6 (95\% CI: [155.1, 209.3]) for Incels.
These values are significantly higher than in Reddit, where there are, on average,
37.3 posts per user (95\% CI: [36.2, 38.5]) in the TD community, and
19.8 (95\% CI: [18.2, 21.5]) in the Incel community. 

Comparing the number of posts per user on Reddit in general (blue line) with users we managed to match (dashed blue line), we find that matched users are more active than users in general.
In Reddit, matched users had an average of 
127.3 posts (95\% CI: [120.2, 135]) in the TD community,
and 319.9 posts (95\% CI: [264.7, 381.8]) in the Incel community, 
significantly higher than the average user in Reddit in each community (reported above).

\xhdr{Matched comparisons} 
The above analysis suggests that users who migrated were more active than average on Reddit, which could lead to an increase in \emph{relative} activity due to self-selection. 
To further investigate this, we compare, for each matched user, the change in number of posts before and after the migration. 
More specifically, we analyze the log-ratio of posts before \vs\ after the migration for each matched user, defined as
$\log_2 \frac{\# \text{ posts after}}{\# \text{ posts before}}$.
Note that this metric provides an intuitive interpretation of the change in activity for a user: if the numbers of posts before and after the migration are the same, the log-ratio will be $0$; if the user posted twice as much, it will be $1$; and if the user posted half as much, $-1$.

\begin{figure}[t]
\centering
\includegraphics[width=\linewidth]{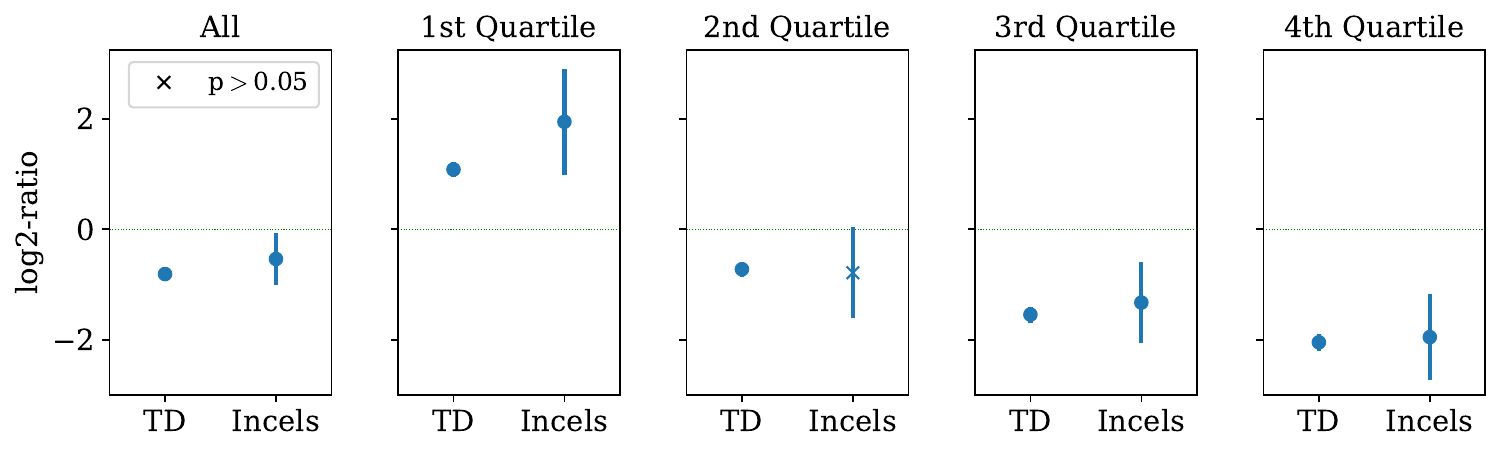}
\caption{
\textbf{User-level change in number of posts:}
Mean log-ratios between the number of posts before and after the migration for each user.
In the first column, the mean is calculated for all users, while for the last four, we stratify users according to their level of pre-migration activity.
The horizontal line depicts the scenario where the number of posts remained the same (log-ratio = 0). Error bars represent 95\% CIs.
}
\label{fig:matched_ratio}
\end{figure}

In \Figref{fig:matched_ratio}, we depict the mean value of the log-ratios for all users in the first column and, in the next four columns, for users stratified by their activity in the pre-migration period.
We divide users in quartiles according to how much they posted in the pre-migration period\footnote{For the TD community, the quartile ranges for the number of posts before the migration were
$Q1=[1,7)$, $Q2=[7,27)$, $Q3=[27,101)$, $Q4=[101,\infty)$; for Incels,
$Q1=[1,19)$, $Q2=[19,116)$, $Q3=[116,398)$, $Q4=[398,\infty)$.}
and then report the mean for each quartile.

Considering the complete set of matched users (first column of \Figref{fig:matched_ratio}), we find that the mean activity log-ratios are significantly smaller than zero for both communities:
-0.81 (95\% CI: [-0.86, -0.75]) for the TD community and
-0.53 (95\% CI: [-0.96, -0.10]) for the Incel community.
This result provides further evidence for the self-selection hypothesis: not only did we find the group of matched users to be more active, but, within this group, activity has \textit{decreased}. 

Analyzing the users stratified by their activity (in the last four columns of \Figref{fig:matched_ratio}), we find that this decrease in activity is stronger for users who were the most active in the pre-migration period.
The mean log-ratios for each quartile in TD are, respectively,
$\mu_{Q1}$ = 1.1, 
$\mu_{Q2}$ = $-$0.7, 
$\mu_{Q3}$ = $-$1.5,
and $\mu_{Q4}$ = $-$2.0.
This shows that users in the least active quartile (Q1) became around twice ($2^{1.1}$) as active, while those in the most active quartile (Q4) decreased their activity to around one-quarter ($2^{-2.0}$).
For the Incel community, we observe a similar pattern, with mean log-ratios of
$\mu_{Q1}$~=~1.9, 
$\mu_{Q2}$~=~$-$0.8,
$\mu_{Q3}$ = $-$1.3, 
and $\mu_{Q4}$~=~$-$1.9.
Overall, these findings mitigate the concern that a core group of extremely dedicated users was ``ignited'' by the migration.

\subsection{Take-aways} 
Our analysis suggests that community-level moderation measures significantly hamper activity and growth in the communities we study.
For both communities, there was a substantial decrease in the number of newcomers, active users, and posts after the moderation measure. 
Yet, this tells only part of the story: we also find an increase in the \emph{relative} activity for both communities: per user, substantially more daily posts occurred on the fringe websites. 

A closer look into \textit{user-level} indicates that this relative increase in activity is due to self-selection, rather than an increase in user activity post-migration.
Not only do we find that users we managed to match were more active on Reddit before the migration, but they even \emph{reduced} their overall activity after they went to the new platform.

\begin{figure*}
\centering
\includegraphics[width=0.99\textwidth]{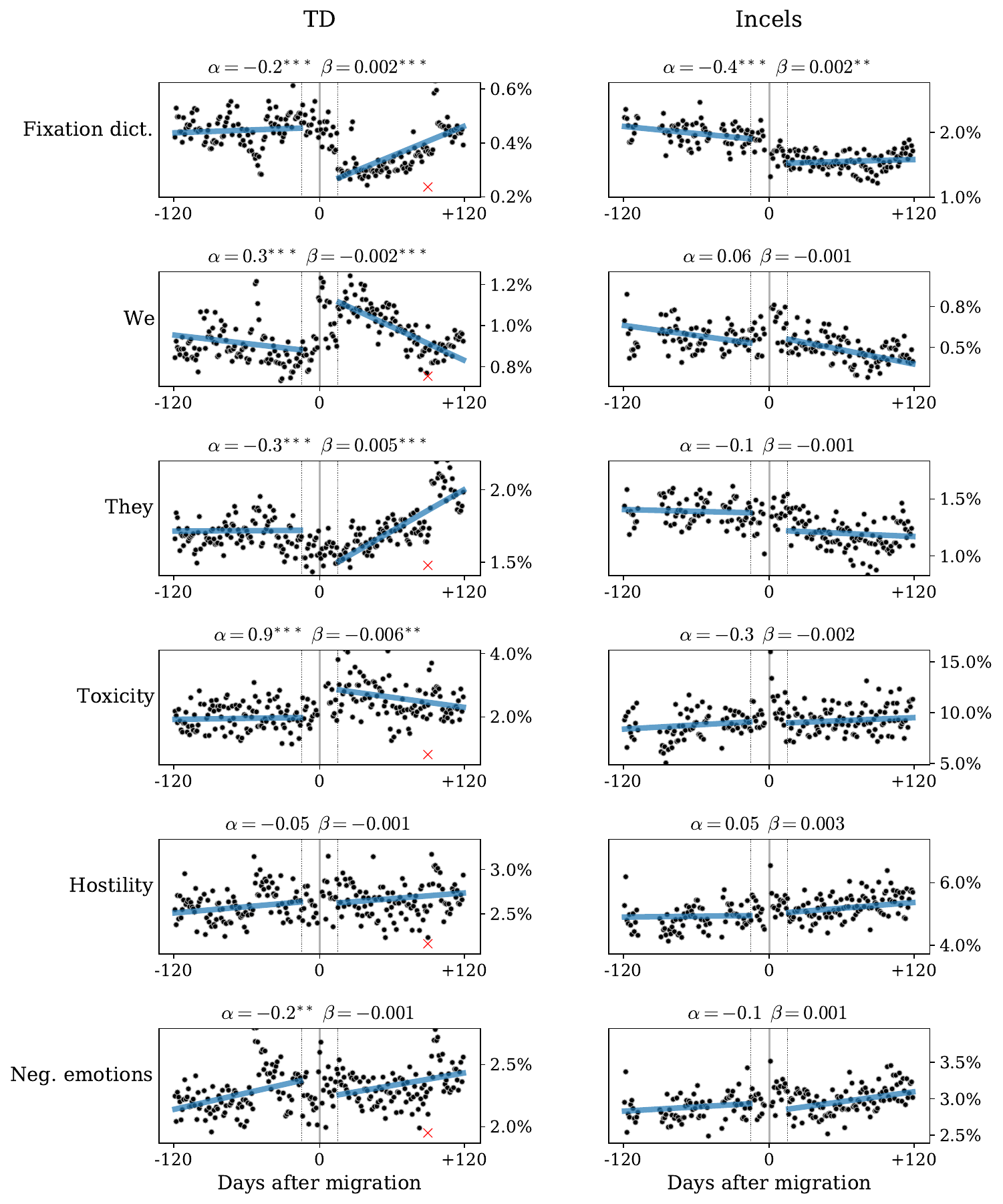}
\caption{
\textbf{Content signals:} 
Daily content-related statistics for the TD community (left) and the Incel community (right) 120 days before and after migrations.
For the \emph{Fixation Dictionary} and the LIWC-related metrics, black dots depict, for each day, the percentage of words belonging to each word category.
For \emph{Toxicity}, they depict the daily percentage of posts with toxicity scores higher than 0.80. 
For \emph{Toxicity}, \emph{Negative Emotions}, and \emph{Hostility}, we limit our analysis to posts that contain at least one word in our fixation dictionaries.
We again show the output of our model as solid blue lines, and the coefficients related to the moderation measure ($\alpha$ and $\beta$) on top of each plot (marking those for which $p<$ 0.001, 0.01, and 0.05  with ***, **, and *, respectively). 
For the TD community, we mark the killing of George Floyd, with a red cross ($\color{red} \times$) close to the x-axis.
}
\label{fig:rdd_tox}
\end{figure*}

\section{Changes in content}

In this section, we use the signals described in \Secref{sec:methods-content} to analyze whether the communities and their users became more toxic and ideologically radical following the migrations. 
We again analyze community- and user-level trends separately.

\subsection{Community-level trends}
\label{sec:content-comlev}
To study community-level trends, we use a regression discontinuity design similar to Equation \eqref{eq:linear}; however, we add an extra term to control for changes in length associated with the migration.%
\footnote{We find significant changes in the average post length pre- \vs post-migration: 131.7 \vs 118.0 for Incels, and  129.2 \vs 141.2 for TD.}
The model now takes on this form:
\begin{equation}
\label{eq:len}
 y_t = \alpha_0 + \beta_0 t + \alpha i_t  + \beta i_t t + \gamma l_t,
\end{equation}
\noindent
where $l_t$ represents the median length of posts (in characters) on day~$t$. 
We add this covariate to ensure that changes in the intercept ($\alpha$) and the slope ($\beta$) following the intervention are not confounded by changes in the way people post on the new platform (e.g., longer posts).
Note that a consequence of this added term is that when we plot the number of posts (on the $y$-axis) per day (on the $x$-axis), we no longer get a straight line since changes in the median length (which varies with time) may impact the outcome of the regression.
Thus, for the plots, we fix the value of the length as the average value through the entire period in order to isolate the effect of the intervention and simulate that there is no length change.
For descriptions of the other coefficients, see Equation~\eqref{eq:linear}.
Again, all coefficients for the regression analysis, along with confidence intervals, are shown in \Tabref{tab:rdd}.

\xhdr{Fixation dictionary} We begin by inspecting the prevalence of the fixation dictionary terms over time, as depicted in the first column of \Figref{fig:rdd_tox}. 
For the TD community, we observe a significant drop of $\alpha=-0.2$ percentage points in the usage of terms in the fixation dictionary
($-44\%$ of the \emph{MVBI}).
For the Incel community, following the intervention, we also see a decrease of around $\alpha~=~-0.4$ percentage points in the usage of words in the fixation dictionary
($-21\%$ of the \emph{MVBI}).
In both cases, we observe a positive increase in the trend after the intervention ($\beta = 0.002$ for both communities). 

\xhdr{Fixation-related signals} Next, we study changes in \emph{Toxicity}, \emph{Negative Emotions}, and \emph{Hostility}. 
We limit this analysis to the set of posts containing at least one word in the fixation dictionary (see \Tabref{tab:fixation}) since we are particularly interested in how the communities are talking about their objects of fixation.
We consider a comment to be toxic if it has a toxicity score above 80\% and calculate, for each day, the fraction of toxic posts.
This threshold has been used as a default in other papers~\cite{zannettou2020measuring} and production-ready applications that use the API~\cite{vox_media_coral_2020}.
For the other LIWC-based metrics, we calculate the proportion of words in the specific dictionaries used per day.

The second column in \Figref{fig:rdd_tox} shows the changes in the percentage of toxic posts for both communities.
For the Incel community, we find no significant change following the interventions.
For TD, there is a significant increase right after the intervention of around $\alpha~=~0.9$ more toxic posts containing the fixation dictionary
($42\%$ of the \emph{MVBI}).
However, we see a significant decreasing trend of around $\beta~=~-0.006$ fewer toxic posts containing words in the fixation dictionary per day. 
This decrease in the overall trend does not necessarily mean that the average percentage of toxic posts will return to the pre-migration levels. 
After the sharp increase in toxicity following the moderation measure, the daily toxicity levels may settle at a new baseline higher than pre-migration values.

The third and fourth columns in \Figref{fig:rdd_tox} depict changes in Negative Emotions and Hostility, respectively.
We find that in most cases these two metrics experience a \emph{decrease} in the intercept following community level interventions, although effects are not always significant ($p >$ 0.05).

\xhdr{Pronoun usage} In the fifth and sixth columns of \Figref{fig:rdd_tox}, we report the usage of two types of personal pronouns: first-person plural pronouns (\eg, ``we,'' ``us,'' ``our'') and third-person plural pronouns (\eg, ``they,'' ``their'').
For the Incel community, we see no significant change in the usage of either type of pronoun following the migration.
For TD, however, there are interesting changes in their usage. 
For first person plural pronouns, following the intervention, we find a significant increase in usage of around $\alpha=0.3$ percentage points ($33\%$ of the \emph{MVBI}), and a significant \emph{decrease} in the slope, $\beta = -0.002$.
For third-person plural pronouns, we find the opposite. Following the intervention,
 we find a significant \textit{decrease} of $\alpha=-0.3$ percentage points ($-18\%$ of the \emph{MVBI}), followed by a significant increase in the trend, $\beta= 0.005$.

First-person plural pronouns capture group identification and third-person plural pronouns have been associated with extremism~\cite{pennebaker2007computerized, grover_detecting_2019, cohen_detecting_2014}. 
Thus, for the TD community, the intervention seems to have transiently increased group identification immediately after the ban, and later, attention seems to have shifted to the outgroup. 
The reduced focus on the outgroup following the community intervention could also be related to the way words in the \textit{fixation dictionary} were used after migration. 
There too, we observe a similar pattern: a sharp drop followed by a gradual increase in usage.

Overall, these findings suggest that the community migrations heterogeneously impacted the communities at hand. 
While not much changed for the Incel community, we find that for TD, there were significant increases in signals related to both the fixation warning behavior (\emph{Toxicity}) and the group identification warning behavior (both first- and third-person plural pronouns).
Again, here a potential confound is the death of George Floyd on 25 May 2020, which impacted user activity (see \Figref{fig:trends}) and coincides with increases in some of the metrics studied (\eg, third-person plural pronouns).
By repeating the analysis for TD excluding the period after 24 May 2020,  we still find that these changes hold. 

\subsection{User-level trends}

Similar to our content-level analysis, the reasons behind the increase in some of the signals related to online radicalization are important.
Here, again, it could be that the subset of users who migrated to the fringe platform was more radical to begin with \emph{or} that the users became more radical after the migration.
Thus, it is important to analyze changes at the user level.
Luckily, the sample of matched users gives us the opportunity to control for self-selection since we can measure, \eg, the percentage of toxic posts before \vs\ after the migration for the same group of matched users.

\begin{figure}[t]
 \centering
 \includegraphics[width=\linewidth]{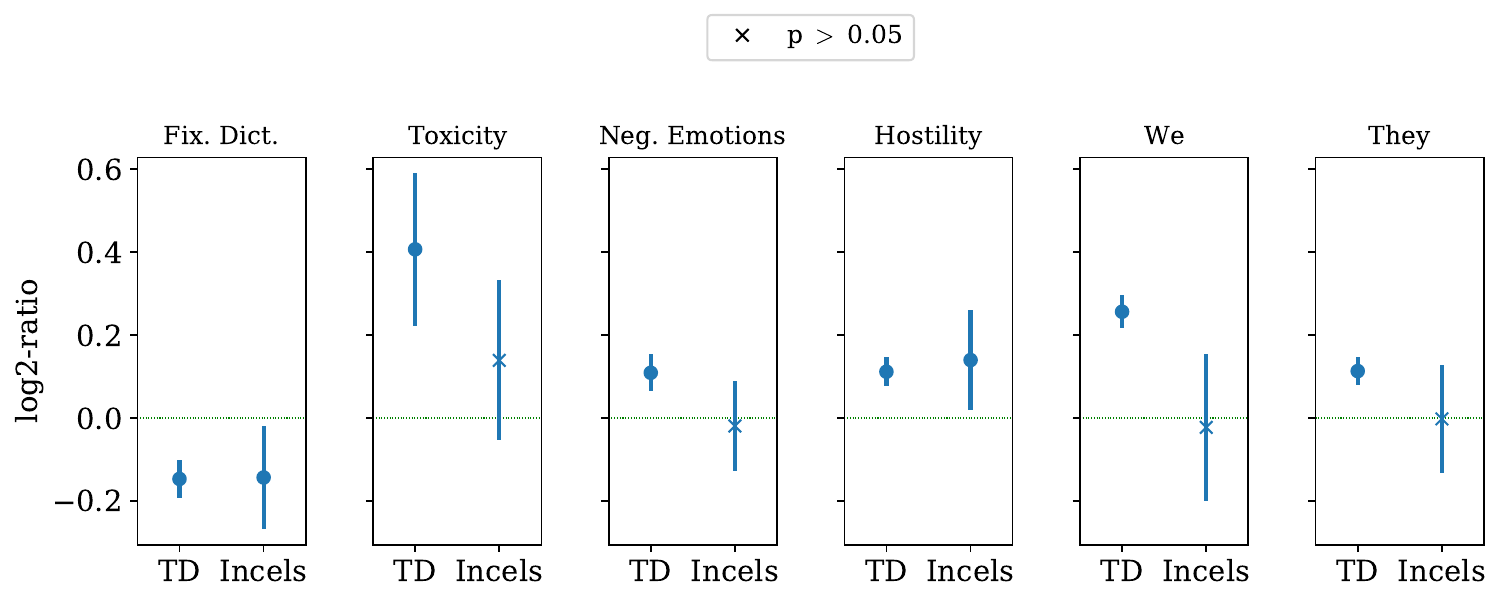}
 \caption{
\textbf{ User-level change in content:}
 We depict the mean user-level log-ratio for each of the content-related signals studied.
 A green horizontal line depicts the scenario of no change (log-ratio = 0).
 Error bars represent 95\% CIs.
 }
 \label{fig:matched_comp_cont}
\end{figure}

\xhdr{Matched comparison}
To disentangle self-selection from user-level increases following the migration, we compare changes  in each of the signals for the set of matched users.
We calculate, for each user, the fraction of toxic posts (\emph{Toxicity} higher than $0.8$) and the percentage of words used in each of the defined categories (\emph{Hostility}, \emph{We}, \etc) both before and after the migration.
Then, similar to \Figref{fig:matched_ratio}, we compare the log-ratio between the signals associated with each user \emph{before} and \emph{after} the migration.
However, here, calculating the log-ratio may involve dividing by 0, \eg, for a user who posted no toxic posts before the migration and 2 after.
Thus, for each individual signal, we limit our analysis to users with positive values for that signal before and after the migration. 
Therefore, when comparing the changes in toxic posts, we consider only users with at least one toxic post before and one toxic post after the migration.
Similarly, for the LIWC-related signals, we consider only users who used words in the given category at least once before and at least once after the migration.
We report the mean log-ratio across  matched users for each signal in \Figref{fig:matched_comp_cont}.

For the TD community, we again observe significant increases for \emph{Toxicity} ($\mu=0.41$,  which represents an increase of around 32\% since $2^{0.41} \approx 1.32$), \emph{We} ($\mu=0.26$, $20\%$ increase), and \emph{They} ($\mu=0.11$, $8\%$ increase).
This suggests that the increases previously observed were not caused merely by self-selection.
For the Incel community, there were non-significant increases in \emph{Toxicity} ($0.14$,  $10\%$ increase) and small non-significant decreases in the usage of both pronoun-related categories. 
For both communities, we again significant decreases in the usage of words in the fixation dictionary ($\mu=0.15$ for TD and $\mu=0.14$ for Incels, around $11\%$ increase in both cases).
We also find significant increases in signals that we did not observe in the community-level analysis.
Namely, for both communities we find significant increases in \emph{Hostility} ($8\%$ increase for TD and $10\%$ for Incels), 
and for TD, we find a significant increase in \textit{Negative Emotions} ($8\%$ increase).

\begin{figure*}[t]
 \centering
 \includegraphics[width=\textwidth]{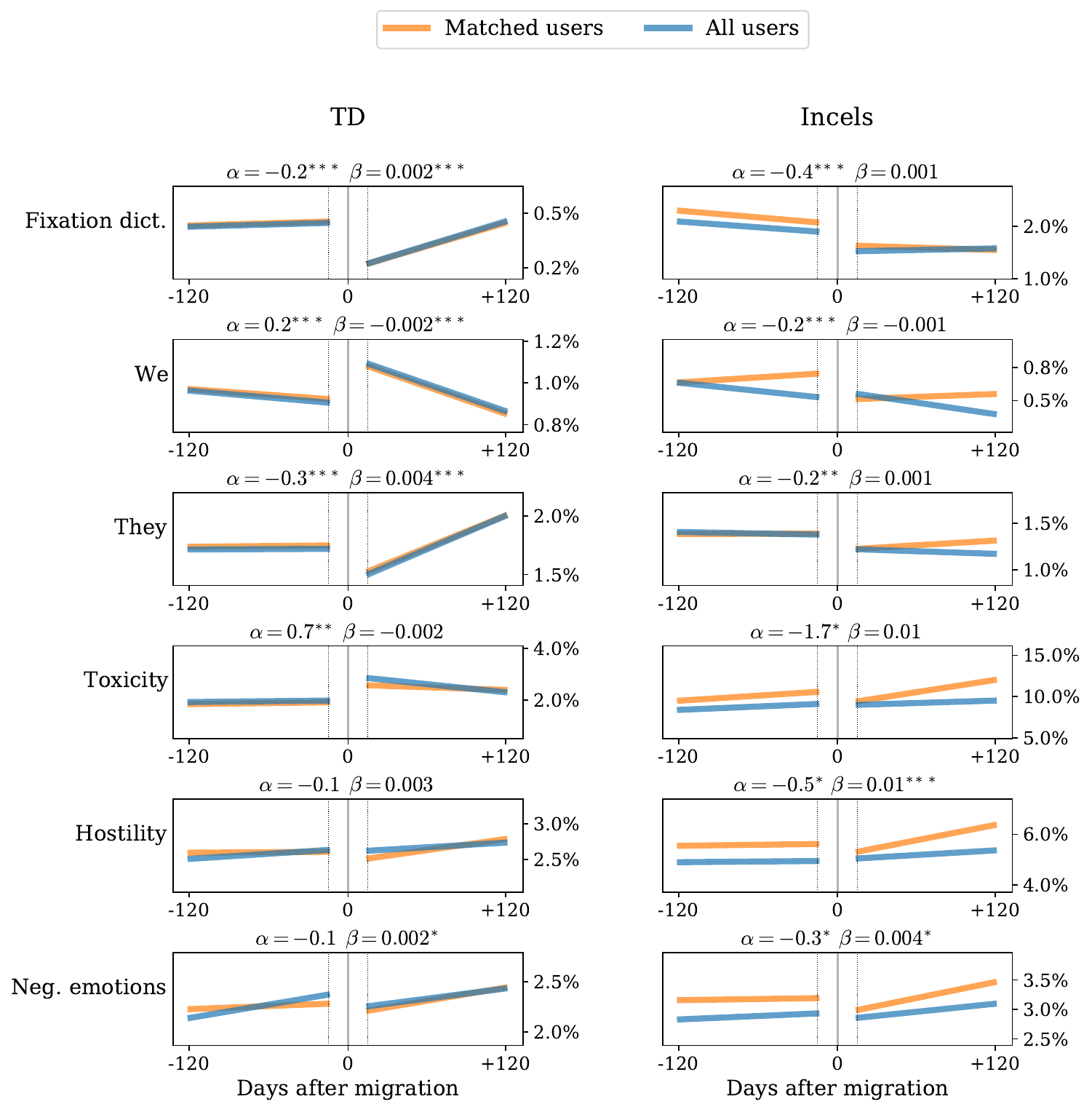}
 \caption{
  \textbf{Daily content signals for matched users:}
  We repeat the same analysis from \Secref{sec:content-comlev} considering the sample of matched users. We show the regression lines considering all users (in blue) and only matched users (in orange). 
  Above each plot, we show the coefficients related to the moderation ($\alpha$ and $\beta$) for the model considering only matched users. 
  For additional details, see \Figref{fig:rdd_tox}.
 }
 \label{fig:rdd_toxu}
\end{figure*}

\xhdr{Regression discontinuity analysis} 
The previous analysis indicates that there were significant changes in the radicalization-related signals for the matched sample, some of which we did not observe in the community-level analysis.
To better understand the matched sample, and the differences between the results at the community-level and the user-level, we repeat the regression discontinuity analysis done for the signals of interest using only posts from the matched user sample. 
We use exactly the same model as in Equation \eqref{eq:len}, changing only the data: \emph{there} we used all posts by all users, \emph{here} we use all posts by matched users. 
In \Figref{fig:rdd_toxu}, we plot the regression lines for the analysis done with all users in blue and for matched users in orange.
Coefficients along with confidence intervals are again presented in \Tabref{tab:rdd}.

For several signals, the results in this reduced sample are very similar to the previous analysis. 
For example, for TD, we have almost exactly the same coefficients for the usage of the fixation dictionary ($\alpha=-0.2$, and $\beta=0.002$) and of third-person plural pronouns   ($\alpha=-0.3$, and $\beta=0.004$).
Yet, for some of the signals, we do find significant differences following the community migrations. 
More specifically, for TD community, following the migration, we find significant increases in the trends for Negative Emotions ($\beta = 0.002$) and we find no significant \textit{decrease} in the trend for the Toxicity signal (which used to be the case).
Additionally, for Incels, we find significant increases in the trend for \emph{Negative Emotions} ($\beta = 0.004$) and \emph{Hostility} ($\beta = 0.01$).

Overall, this analysis confirms the results previously discussed in \Figref{fig:matched_comp_cont} and suggests that users in the matched sample were impacted by the community-level intervention.
This is different from what we observed when looking at activity levels. 
There, when we zoomed in on matched users, we found that they had \textit{decreased} their activity (even though the number of posts per user grew).
Here, on the contrary, we find that these users seem to have become more radical.

\subsection{Take-aways} 

Altogether,  our analysis shows that, for TD, community-level interventions and the migrations that ensued are associated with significant increases in radicalization-related signals.
A closer look at the matched user sample indicates that these increases were not merely due to self-selection, since we also observe  significant user-level increases. 
Furthermore, analyzing the matched sample, we find that the migration may have impacted these users more substantially, since the differences for them are more substantial. 

A second important result of our content-level analysis is that communities were heterogeneously impacted. 
When comparing how the \emph{activity} in the two communities changed (\Secref{sec:activity}), we found the same patterns overall; whereas, when comparing how the \emph{content} changed, we found rather distinct behaviors across the two communities. 
Unlike the TD community, for Incels, there were often \emph{decreases} in signals related to radicalization following community migration.

\section{Discussion \& conclusion}
\label{sec:discussion}

Our work paints a nuanced portrait of the benefits and possible backlashes of community-level interventions. 
On the one hand, we found that the interventions were effective in decreasing activity and the capacity of the community to attract newcomers. 
Moreover, we found evidence that \textit{relative} increase in activity (i.e., fewer users posting more) is likely due to self-selection: the users who migrated to the new community were more active to begin with.
On the other hand, we found significant increases in radicalization\hyp related signals for one of the communities studied (TD), even when controlling for self-selection. 
In fact, these increases were even more substantial for the set of matched users studied.

An interesting angle to consider the changes observed in communities pre\hyp{} \vs pos\hyp{} migration, is through the lens of  characteristics and affordances of large online platforms such as Reddit, YouTube and Facebook.
According to Gillespie~\cite{gillespie_custodians_2018}, online platforms differ from traditional media outlets in that they provide the means of distribution, but not the content (which is user generated). 
Moreover, a key component of online platforms is that they moderate and gatekeep content, despite  best efforts to present themselves as neutral ``facilitators.''

In that context, the migration of online communities from mainstream platforms to fringe, alternative websites provoke shifts associated with how content is distributed and moderated, two important roles of online platforms.
The decrease in activity observed after communities migrated emphasizes the \emph{power of the distribution} of online platforms such as Reddit. 
Since Reddit has thousands of highly popular subreddits, toxic communities inhabiting the platform are easily discoverable, and consuming the content they produce is convenient.
Using a similar line of reasoning, the increase in toxicity observed when members of r/The\_Donald migrated out of the subreddit can be associated with the \emph{power of the moderation} of online platforms.
Toxicity may be understood as a proxy for content that would likely clash with Reddit's content policy\footnote{\url{https://www.reddit.com/r/reddit.com/wiki/revisions/contentpolicy}}. 
Therefore, the rise in toxicity following the ban can be understood as a consequence of the removal of platform moderation.

Overall, our results strengthen the hypothesis that platforms are largely responsible for our online information ecosystem~\cite{gillespie_custodians_2018}. Besides determining what kinds of content flourishes~\cite{munger2020right}, platforms allow communities to exploit their affordances to recruit new members, and are able to influence the content being posted in toxic communities.
In the remainder of this section, we discuss the implications of these results for platforms and future research, as well as the limitations of our study.

\subsection{Limitations and future work}

\xhdr{Communities} 
Our work focuses on two communities: TD and Incels.
However, Reddit has sanctioned many other communities that may have migrated to new fringe websites.
The implications of such sanctions for migration may differ based on the specifics of each community.
That said, the communities we study are among the most prominently sanctioned subreddits, and our analysis provides early insight into the consequences of such sanctions.
In the future, similar analysis on other sanctioned communities would help disentangle how contextual factors including community size, topic, and the design of the alternative platform may affect migration patterns.

\xhdr{Migrations and dispersion} 
We consider the effects of migration to only one fringe website per each of the sanctioned communities we study. 
In both cases, the migrations to the websites we analyze were officially endorsed by the subreddits' moderators, and, for r/The\_Donald, the subreddit promoted the migration to the new site while it could. 
However, users may have migrated to other platforms as well. 
For example, on Reddit, after r/Incels was banned, an old subreddit called r/Braincels reportedly became popular (until eventually being banned too). 
Also more broadly, some community-level interventions may not result in ``successful'' coordinated migration.
Rather, users can be dispersed through a variety of other platforms (\eg, Gab, 4chan, Parler, \etc).
Studying what happens in these cases is an important direction to completely understand the impact of deplatforming communities. For example, one could try to measure the activity boost experienced in each of these platforms whenever a toxic community in a mainstream platform (\eg r/The\_Donald) gets banned. 
A  challenge here would be to obtain data for a variety of fringe communities and to control for other confounders, such as geopolitical events.

\xhdr{Confounders} 
The responsiveness of these communities to real-world events creates confounders.
This is particularly true for the TD community, where we found significant changes in the content- and activity-related signals in reaction to the killing of George Floyd.
While our quasi-experimental research design controls for linear trends, sudden bursts in content-related signals can partially impact our results.
Controlling for these trends is hard since the reaction of these communities to real world changes is inherently linked to the harms they pose to society.
However, in our specific case, we find that the effects observed held even when limiting the period of the regression discontinuity analysis to before the event (i.e., George Floyd's killing).
Another possible set of confounders are changes to rules and moderation actions that could have changed pre\hyp{} \vs post\hyp{}intervention. 
Although we did not explicitly incorporate these changes into our analysis, we carefully analyzed the set of rules before (Incels: ~\cite{incelrules1}, TD:~\cite{tdrules1}) and after (Incels: ~\cite{incelrules2}, TD:~\cite{tdrules2}) the migration and did not find any substantial changes.

\xhdr{Matched Users}
Another limitation of the work at hand is that user-level analyses are made on a set of users matched according to their usernames. 
These users tend to be more active than the average user (\cf \Figref{fig:matched_summary}), and may differ from users who migrated and did not change their username.
Although important, we argue that this bias is not impactful to the external validity of our results.
The main purpose of looking at matched users is to distinguish between behavior change and self-selection.
When studying changes in activity, analyzing matched users provides us with the useful insight that, although the number of posts per user increases after migration (\cf \Figref{fig:trends}), on a  user-level this is not the case (\cf \Figref{fig:matched_ratio}).
For the sample bias to be an issue here, reverting or weakening the results, it would be necessary that users who migrated and did not keep the same username became more active after the ban while those who kept the same username did not, which is unlikely.
When studying changes in the content, we find that community-level trends on the set of matched users are very similar to community-level changes considering all users (\cf \Figref{fig:rdd_toxu}), weakening concerns that there would be a strong difference between the nature of the content posted by these users.
An interesting direction to further understand these matched users (and explore user-level trends) would be to additionally analyze users with known usernames pre\hyp{} \vs post\hyp{} ban in other subreddits. 

\xhdr{Mapping signals to externalities}
Our analysis relies on user activity and signals derived from user-generated content to analyze online toxic communities.
Our main result suggests that community-level interventions may involve a trade-off: less activity at the expense of a more radical community elsewhere.
Yet, the relationship between these activity- and content-related signals from toxic online communities and their real-world harms is still fuzzy. 
It is unclear, for instance, whether a reduction of 50\% in posting activity where each user is 10\% more ``toxic'' is desirable or not.
While such a fine-grained assessment of the consequences of a moderation intervention is out of the scope of this paper, further study of the causal links between toxicity, user activity, and real-world harm is an important research direction to improve the quality of moderation decisions.

\subsection{Implications for online platforms}
Our analysis of migration dynamics highlights that community-wide moderation interventions do not happen in a vacuum. 
When platforms sanction an entire community, as opposed to taking user-level actions, communities may migrate \textit{en gros} to a different platform. 
Platforms have difficult decisions to make: they need to consider the effects of community-wide sanctions not only on their own backyard, but on other online and offline spaces as well. 
Our results suggest that there may be a trade-off associated with this decision: banning a community from a mainstream platform may come at the expense of a smaller but more extreme community elsewhere.
However, this take-away should be handled with nuance, since our work is limited to two communities, and since the increase in toxicity was only observed in one of the two communities. 

Nevertheless, a practical implication that follows from our results is that, given that a community eventually gets banned, the time the said community was allowed to flourish in a mainstream platform may increase its potential for harm post-banning.
The reasoning is simple: since community growth is halted by the deplatforming, the earlier the community is banned, the fewer members a possible spin-off community would have.
In that context, if banning is a commonly used practice in a given platform, it is advantageous to employ the measure proactively rather than reactively.

Lastly, the methodological framework we use in this paper may also be used in other contexts and platforms to evaluate the effectiveness of moderation interventions.
Platforms have at their disposal abundant data that can help further clarify the trade-offs we discussed here.
We hope that extensions of this work will yield more precise guidelines on how to handle problematic online communities.

\section*{Acknowledgements}
Manoel Horta Ribeiro is supported by a Facebook Fellowship Award.
Jeremy Blackburn is supported by NSF grants CNS-2114411 and IIS-2046590.
Gianluca Stringhini is supported by NSF grants CNS-1942610 and CNS-2114407.
Emiliano De Cristofaro is supported by the UK's National Research Centre on Privacy, Harm Reduction, and Adversarial Influence Online (REPHRAIN, UKRI grant: EP/V011189/1).
Robert West is partly supported by a grant from the EPFL/UNIL Collaborative Research on Science and Society (CROSS) Program, the Swiss National Science Foundation (grant 200021\_185043), the European Union (TAILOR, grant 952215), 
and gifts from Google, Facebook, and Microsoft.

\begin{table*}
 \centering
 \caption{Coefficients for all regression discontinuity analyses done throughout the paper, including 95\% confidence intervals. 
 Coefficients for which $p<$ 0.001, 0.01, and 0.05 are marked with ***, **, and *, respectively.  
 The value $[10^{-3}]$ at the beginning of a cell indicates that the value of the cell as well as the confidence intervals presented should be multiplied by $10^{-3}$. 
 This may cause slight differences in the numbers in this table and the ones presented in the plots, since here we present the results at higher precision.
 Note that this table contains the regression results for three different analysis carried out throughout the paper and depicted in \Figref{fig:trends}, \Figref{fig:rdd_tox}, and \Figref{fig:rdd_toxu}. 
 For presentation reasons, we omit the confidence intervals for the intercept across the whole period ($\alpha_0$), which is significant ($p<0.001$) across all of the models.}
\label{tab:rdd}
\input{images/big_fat_table}
\end{table*}

%% file: images/overview_datasets.tex
\begin{tabular}{l|r|r|r|r}
\toprule
\multicolumn{1}{l}{\textbf{Platform}} & \multicolumn{1}{l}{\textbf{Community}} & \multicolumn{1}{l}{\textbf{Submissions}} & \multicolumn{1}{l}{\textbf{Comments}} & \multicolumn{1}{l}{\textbf{Users}} \\ \midrule
\multirow{2}{*}{Reddit} & \texttt{/r/Incels} & 17,403 & 340,650 & 18,088 \\
 & \texttt{/r/The\_Donald} &  251,090 & 2,703,615 & 80,002 \\ \midrule
\multirow{2}{*}{Websites} & \texttt{Incels.co} & 25,138 & 385,765 & 2,270 \\
 & \texttt{thedonald.win} & 280,156 & 2,390,641 & 38,510 \\ \bottomrule
\end{tabular}%

%% file: images/fix_dict.tex
\begin{tabular}{p{0.075\linewidth}|p{0.8\linewidth}}
\toprule
Incels & 
female(s)
normie(s)
chad(s)
virgin
whore(s)
girl(s)
rope
gf
girlfriend
women
beta
cunt
suicide
pussy
woman
bitch(es)
cuck(s)
feminism \\ \midrule
TD & 
trans
commie
dem(s)
democrat(s)
deep
communist
diversity
leftist
communism
antifa
socialist
left
socialism
libs
gender\\ \bottomrule
\end{tabular}

%% file: images/big_fat_table.tex
\vspace{3mm}
{
\subcaption{Community-level activity (\Figref{fig:trends})}
\tiny
\begin{tabular}{lllllll}
\toprule
          &                 &     $\alpha_0$ &                    $\beta_0$ &                         $\alpha$ &                      $\beta$ &   $R^2$ \\
Venue & Statistic &                &                              &                                  &                              &         \\
\midrule
TD & \#newcomers &  $214.7^{***}$ &        $-0.5^{} (-1.3, 0.2)$ &     $-77.8^{**} (-127.9, -27.7)$ &         $0.5^{} (-0.5, 1.4)$ &  $0.24$ \\
          & \#posts &  $26650^{***}$ &       $8.8^{} (-17.9, 35.5)$ &  $-14416^{***} (-16947, -11886)$ &  $120.6^{***} (83.9, 157.2)$ &  $0.41$ \\
          & \#users &   $7593^{***}$ &           $4^{} (-0.5, 8.5)$ &     $-4774^{***} (-5184, -4365)$ &     $13.7^{***} (8.1, 19.3)$ &  $0.85$ \\
          & \#posts/\#users &    $3.5^{***}$ &  $-0.001^{} (-0.003, 0.001)$ &           $1.1^{***} (0.9, 1.3)$ &  $0.009^{***} (0.006, 0.01)$ &  $0.85$ \\
Incels & \#newcomers &  $224.9^{***}$ &       $0.9^{***} (0.5, 1.4)$ &  $-215.4^{***} (-250.3, -180.5)$ &    $-0.9^{***} (-1.3, -0.4)$ &  $0.84$ \\
          & \#posts &   $4840^{***}$ &      $19.5^{***} (14.1, 25)$ &     $-2651^{***} (-3098, -2203)$ &         $1.7^{} (-4.8, 8.2)$ &  $0.55$ \\
          & \#users &  $960.1^{***}$ &       $3.9^{***} (2.9, 4.9)$ &  $-777.4^{***} (-850.2, -704.6)$ &        $-3^{***} (-4, -2.1)$ &  $0.93$ \\
          & \#posts/\#users &      $5^{***}$ &  $-0.001^{} (-0.003, 0.003)$ &           $7.4^{***} (6.6, 8.3)$ &    $0.04^{***} (0.02, 0.05)$ &  $0.92$ \\
\bottomrule
\end{tabular}

}
\vspace{8.5mm}
{
\subcaption{Community-level content (\Figref{fig:rdd_tox})}

\tiny
\begin{tabular}{lllllll}
\toprule
          &      &                $\alpha_0$ &                           $\beta_0$ &                   $\alpha$ &                              $\beta$ &   $R^2$ \\
Venue & Statistic &                           &                                     &                            &                                      &         \\
\midrule
TD & Fix. Dict &    $0.6^{***} $ &      $[10^{-3}] 0.2^{} (-0.2, 0.5)$ &  $-0.2^{***} (-0.3, -0.2)$ &     $[10^{-3}] 1.7^{***} (1.2, 2.2)$ &  $0.53$ \\
          & Toxicity &    $3.1^{***} $ &        $[10^{-3}] 0.5^{} (-2, 3.1)$ &     $0.9^{***} (0.6, 1.3)$ &  $[10^{-3}] -5.9^{**} (-10.1, -1.7)$ &   $0.3$ \\
          & Neg. Emotion &    $2.7^{***} $ &    $[10^{-3}] 2.2^{***} (1.1, 3.3)$ &  $-0.2^{**} (-0.3, -0.06)$ &      $[10^{-3}] -0.5^{} (-2.1, 1.1)$ &  $0.14$ \\
          & Hostility &    $3.3^{***} $ &     $[10^{-3}] 1.2^{} (-0.04, 2.4)$ &    $-0.05^{} (-0.2, 0.09)$ &        $[10^{-3}] -0.1^{} (-2, 1.8)$ &  $0.08$ \\
          & We &    $0.6^{***} $ &  $[10^{-3}] -0.7^{**} (-1.2, -0.2)$ &     $0.3^{***} (0.2, 0.3)$ &  $[10^{-3}] -2.1^{***} (-2.7, -1.4)$ &  $0.46$ \\
          & They &    $1.3^{***} $ &     $[10^{-3}] 0.05^{} (-0.5, 0.6)$ &  $-0.3^{***} (-0.4, -0.2)$ &     $[10^{-3}] 4.8^{***} (3.8, 5.7)$ &  $0.55$ \\
Incels & Fix. Dict &    $1.9^{***} $ &   $[10^{-3}] -1.8^{*} (-3.3, -0.4)$ &  $-0.4^{***} (-0.5, -0.2)$ &      $[10^{-3}] 2.4^{**} (0.9, 3.8)$ &  $0.69$ \\
          & Toxicity &  $11.4^{***}$ &     $[10^{-3}] 6.9^{} (-3.1, 16.8)$ &      $-0.3^{} (-1.2, 0.6)$ &     $[10^{-3}] -1.9^{} (-13.7, 9.9)$ &  $0.13$ \\
          & Neg. Emotion &    $3.2^{***} $ &        $[10^{-3}] 1^{} (-0.7, 2.7)$ &     $-0.1^{} (-0.3, 0.03)$ &       $[10^{-3}] 1.3^{} (-0.4, 3.1)$ &  $0.25$ \\
          & Hostility &    $6.3^{***} $ &        $[10^{-3}] 0.4^{} (-3.1, 4)$ &      $0.05^{} (-0.3, 0.4)$ &       $[10^{-3}] 2.6^{} (-0.8, 6.1)$ &  $0.38$ \\
          & We &    $0.6^{***}$ &       $[10^{-3}] -1^{} (-2.5, 0.4)$ &     $0.06^{} (-0.05, 0.2)$ &      $[10^{-3}] -0.4^{} (-1.3, 0.5)$ &   $0.3$ \\
          & They &    $1.2^{***} $ &     $[10^{-3}] -0.3^{} (-2.6, 2.1)$ &     $-0.1^{} (-0.3, 0.04)$ &      $[10^{-3}] -0.2^{} (-1.6, 1.2)$ &  $0.44$ \\
\bottomrule
\end{tabular}
}
\vspace{8.5mm}
{
\subcaption{Community-level content for matched users (\Figref{fig:rdd_toxu})}

\tiny
\begin{tabular}{lllllll}
\toprule
          &      &                $\alpha_0$ &                          $\beta_0$ &                   $\alpha$ &                              $\beta$ &   $R^2$ \\
Venue & Statistic &                           &                                    &                            &                                      &         \\
\midrule
TD & Fix. Dict &  $0.6^{***} (0.5, 0.8)$ &     $[10^{-3}] 0.2^{} (-0.2, 0.5)$ &  $-0.2^{***} (-0.3, -0.2)$ &     $[10^{-3}] 1.6^{***} (1.1, 2.1)$ &  $0.54$ \\
          & Toxicity &  $3.2^{***} (2.1, 4.3)$ &       $[10^{-3}] 0.6^{} (-2.8, 4)$ &      $0.7^{**} (0.2, 1.2)$ &        $[10^{-3}] -2.4^{} (-9, 4.3)$ &  $0.13$ \\
          & Neg. Emotion &  $2.5^{***} (2.2, 2.8)$ &     $[10^{-3}] 0.5^{} (-0.5, 1.6)$ &    $-0.1^{} (-0.2, 0.006)$ &      $[10^{-3}] 1.7^{*} (0.06, 3.3)$ &  $0.08$ \\
          & Hostility &  $3.1^{***} (2.7, 3.5)$ &     $[10^{-3}] 0.1^{} (-1.4, 1.6)$ &     $-0.1^{} (-0.3, 0.04)$ &       $[10^{-3}] 2.5^{} (-0.001, 5)$ &  $0.06$ \\
          & We &  $0.5^{***} (0.2, 0.8)$ &  $[10^{-3}] -0.6^{} (-1.2, 0.007)$ &     $0.2^{***} (0.2, 0.3)$ &  $[10^{-3}] -2.2^{***} (-2.9, -1.4)$ &  $0.37$ \\
          & They &  $1.6^{***} (1.3, 1.8)$ &     $[10^{-3}] 0.1^{} (-0.6, 0.8)$ &  $-0.3^{***} (-0.4, -0.2)$ &     $[10^{-3}] 4.5^{***} (3.4, 5.6)$ &  $0.44$ \\
Incels & Fix. Dict &  $1.9^{***} (1.6, 2.2)$ &  $[10^{-3}] -2.1^{*} (-3.9, -0.4)$ &  $-0.4^{***} (-0.6, -0.2)$ &       $[10^{-3}] 1.4^{} (-0.9, 3.6)$ &  $0.68$ \\
          & Toxicity &  $11^{***} (7.9, 14.1)$ &           $0.01^{} (-0.008, 0.03)$ &    $-1.7^{*} (-3.2, -0.1)$ &             $0.01^{} (-0.007, 0.04)$ &  $0.08$ \\
          & Neg. Emotion &    $3^{***} (2.6, 3.4)$ &       $[10^{-3}] 0.3^{} (-3, 3.7)$ &   $-0.3^{*} (-0.5, -0.04)$ &       $[10^{-3}] 4.2^{*} (0.6, 7.8)$ &   $0.1$ \\
          & Hostility &    $6^{***} (5.3, 6.7)$ &     $[10^{-3}] 0.7^{} (-3.9, 5.3)$ &   $-0.5^{*} (-0.9, -0.06)$ &      $[10^{-3}] 9.5^{***} (4, 15.1)$ &   $0.2$ \\
          & We &  $0.4^{***} (0.3, 0.6)$ &     $[10^{-3}] 0.6^{} (-0.6, 1.8)$ &  $-0.2^{***} (-0.3, -0.1)$ &        $[10^{-3}] -0.3^{} (-1.5, 1)$ &  $0.44$ \\
          & They &  $0.9^{***} (0.7, 1.2)$ &    $[10^{-3}] 0.04^{} (-1.7, 1.8)$ &  $-0.2^{**} (-0.3, -0.04)$ &       $[10^{-3}] 0.8^{} (-1.2, 2.8)$ &  $0.28$ \\
\bottomrule
\end{tabular}
}